\documentclass[%
 aip,
 amsmath,amssymb,
 reprint,%
]{revtex4-1}

\usepackage{graphicx}
\usepackage{dcolumn}
\usepackage{bm}

\usepackage[utf8]{inputenc}
\usepackage[T1]{fontenc}
\usepackage{mathptmx}
\usepackage{etoolbox}
\usepackage{color}
\makeatletter
\def\@email#1#2{%
 \endgroup
 \patchcmd{\titleblock@produce}
  {\frontmatter@RRAPformat}
  {\frontmatter@RRAPformat{\produce@RRAP{*#1\href{mailto:#2}{#2}}}\frontmatter@RRAPformat}
  {}{}
}%
\makeatother
\begin{document}

\preprint{AIP/123-QED}

\title[Liu et al. Higher-order link prediction via local information]{Higher-order link prediction via local information}
\author{Bo Liu}
\affiliation{ 
Institute of Fundamental and Frontier Sciences, University of Electronic Science and Technology of China, Chengdu, 610054, P. R. China.
}%
\affiliation{ 
Yangtze Delta Region Institute (Huzhou), University of Electronic Science and Technology of China, Huzhou, 313001, P. R. China.
}%
\author{Rongmei Yang}%
\affiliation{ 
Institute of Fundamental and Frontier Sciences, University of Electronic Science and Technology of China, Chengdu, 610054, P. R. China.
}%
\author{Linyuan L\"u*}
\affiliation{ 
Institute of Fundamental and Frontier Sciences, University of Electronic Science and Technology of China, Chengdu, 610054, P. R. China.
}%
\affiliation{ 
Yangtze Delta Region Institute (Huzhou), University of Electronic Science and Technology of China, Huzhou, 313001, P. R. China.
}%
 \email{linyuan.lv@uestc.edu.cn}

\date{\today}

\begin{abstract}
Link prediction has been widely studied as an important research direction. Higher-order link prediction has gained especially significant attention since higher-order networks provide a more accurate description of real-world complex systems. However, higher-order networks contain more complex information than traditional pairwise networks, making the prediction of higher-order links a formidable challenging task. Recently, researchers have discovered that local features have advantages over long-range features in higher-order link prediction. Therefore, it is necessary to develop more efficient and concise higher-order link prediction algorithms based on local features. In this paper, we proposed two similarity metrics via local information, simplicial decomposition weight (SDW) and closed ratio weight (CRW), to predict possible future higher-order interactions (simplices) in simplicial networks. These two algorithms capture local higher-order information at two aspects: simplex decomposition and cliques’ state (closed or open). We tested their performance in eight empirical simplicial networks, and the results show that our proposed metrics outperform other benchmarks in predicting third-order and fourth-order interactions (simplices) in most cases. In addition, we explore the robustness of the proposed algorithms, and the results suggest that the performance of these novel algorithms is advanced under different sizes of training sets.
\end{abstract}

\maketitle

\begin{quotation}
Higher-order networks have received more attention because they can more accurately describe real-world systems. Therefore, it is highly practical to predict higher-order interactions in complex systems. However, existing higher-order link prediction algorithms have limitations and can only forecast third-order interactions. Thus, there is a need to develop a more efficient and effective higher-order link prediction algorithm to address this gap. In this paper, we proposed two novel similarity metrics based on local information, SDW and CRW. We tested their effectiveness by comparing them with five existing algorithms in third-order and fourth-order link prediction on eight empirical networks. In general, the results show that the proposed metrics not only have the best prediction performance in most cases, but also have better robustness.
\end{quotation}

\section{Introduction}
\label{sec:Introduction}
With the fast development of information technology and data explosion, almost complex systems in real life can be abstracted as complex networks of nodes and edges \cite{strogatz2001exploring,boccaletti2006complex}, such as social networks \cite{wasserman1994social}, collaboration networks\cite{newman2001structure} and biological networks\cite{girvan2002community}. Networks come in various complex structures, each with its own unique characteristics. Hence, mining valuable information in networks has always been one of the key scientific problems in network science.

Link prediction has been a hot research topic in network information mining. The task of link prediction is predicting the missing or future links based on the observed data \cite{liben2003link,lu2011link,zhou2021progresses,yijun2022maximum}. Link prediction can not only promote the development of network science and information science but also has a wide range of application prospects. For instance, companies can use link prediction to enhance recommendation systems by forecasting potential connections between users and items, making recommendations more efficient and accurate\cite{huang2005link,li2014recommendation}. Researchers often use the concepts of spurious and missing edges in link prediction to evaluate the reliability of a network and make decisions about its reconfiguration\cite{guimera2009missing}. Link prediction can also be applied to identify potentially pathogenic microRNAs associate with disease, For example, in the case of breast cancer, the top 20 microRNAs that were predicted with this approach have been validated by manual experiments.\cite{turki2017link}.

More recently, there is growing evidence that interactions between agents are often beyond pairwise relationships in many real-world systems \cite{pearcy2014hypergraph,mayfield2017higher,centola2018experimental,milojevic2014principles}. For example, a metabolic reaction is carried out by multiple reactants, an article often involves cooperation between several authors, and in social activities, it is common for multiple individuals to communicate simultaneously. These phenomena promote the formation of higher-order network research \cite{battiston2020networks,battiston2021physics}. Hypergraphs and simplicial networks (complexes) are two common representations of higher-order networks\cite{battiston2020networks}. Hypergraphs encode the interaction between multiple nodes through a node set and hyperedges, which are subsets of the nodeset. Simplicial networks provide an alternative way to represent higher-order interactions using algebraic topology. Simplicial networks are collections of simplices and more constrained than hypergraphs. A simplex requires the presence of all possible subsimplices\cite{bianconi2021higher}. Recently, scientists have begun to explore the structure and dynamics of simplicial networks, and found that simplicial networks provide new insights in percolation\cite{zhao2022higher,wang2022generalized}, synchronization\cite{millan2020explosive,lucas2020multiorder} and social contagion \cite{iacopini2019simplicial,chowdhary2021simplicial}.

An important and challenging problem is predicting higher-order interactions in higher-order networks. Higher-order link prediction can be defined as predicting which higher-order interactions are missing or likely to occur in the future. Recently, higher-order link prediction in simplicial networks has been paid more attention. Benson et al. showed that local information plays a more critical role than long-range information in predicting higher-order links\cite{benson2018simplicial}. Chavan et al. extended the traditional graph embedding algorithm to triangle embedding to predict higher-order links in the network\cite{chavan2020higher}.  However, the existing methods for predicting higher-order links in simplicial networks are relatively rare, and most of them are merely extensions of pairwise link prediction algorithms.

This paper mainly discusses how to use more local information to predict higher-order interactions in simplicial networks. We first proposed two novel higher-order similarity indices based on local information. One index is simplex decomposition weight (SDW), which is based on simplex decomposition, and another index is closed ratio weight (CRW), which is based on the state of cliques. Secondly, we evaluated the performance of the two proposed algorithms on eight empirical networks by making them to predict third-order and fourth-order interactions. This also marks the first prediction of fourth-order interactions. Finally, the robustness of the proposed indices was tested with different proportions of the known data.

In summary, our work has made significant contributions in the following ways: (i) We have proposed two novel indices to describe the similarity of higher-order interactions from different perspectives. (ii) By comparing our proposed indices with existing ones that rely on higher-order local information, we have demonstrated that our indices have distinct advantages and greater robustness.

\section{Methods}
\label{sec:Method}
\subsection{Temporal simplicial network and simplicial closure event}
\begin{figure*}[htb]
\includegraphics{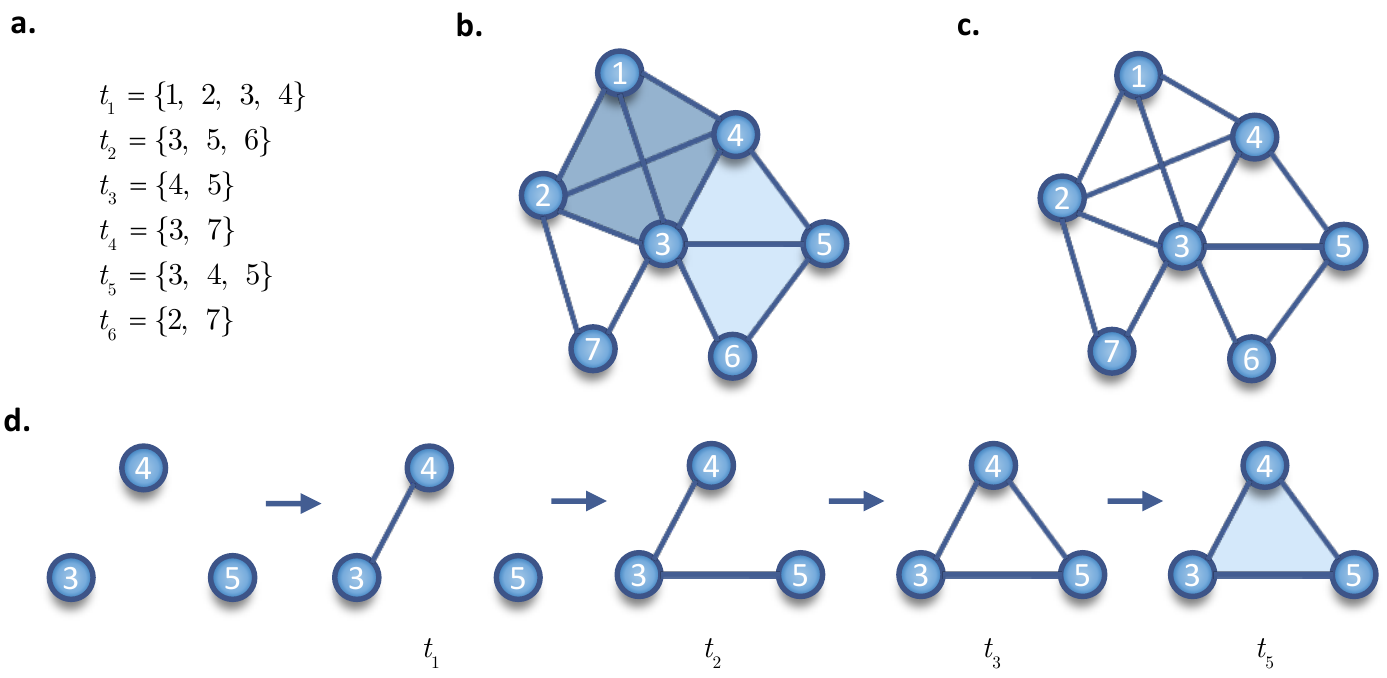}
\caption{\textbf{(Color online) The higher-order interactions data and it's topological structure and simplex closure event.} \textbf{(a)} a set of higher-order interactions data; \textbf{(b)} the visualization of the simplicial network which is composed of higher-order interactions data in (a); \textbf{(c)} the skeleton network of this simplicial network; \textbf{(d)} a simplicial closure event of nodes 3, 4 and 5.}
\label{fig1}
\end{figure*}

In this paper, we have constructed simplicial networks using simplex data with timestamps. Fig.~\ref{fig1}(a) shows the interaction data with timestamps. Considering each interaction as a simplex, we can obtain a simplicial network, as is visualized in Fig.~\ref{fig1}(b). Each simplex in a simplicial network satisfies the closure property, i.e., it contains all sub-simplices that consist of its nodes and are of a lower order than it. For example, if a 3-simplex (3,4,5) exists, one must require that all the simplices \{(3,4,5), (3,4), (3,5), (4,5), (3), (4), (5)\} exist. The closure property is one of the most significant differences between simplicial networks and hypergraphs or the extraction of cliques in pairwise networks. In this study, we focus on higher-order link prediction in temporal simplicial networks. For a clearer understanding, we define a $k$-simplex to be composed of $k$ nodes (another definition is $k+1$ nodes in some other studies). In previous studies, researchers have reported if there is an all-order subset of $k$-simplex between $k$ nodes, they tend to form a $k$-simplex\cite{benson2018simplicial}. The structure with $k$ nodes and an all-order subset of a $k$-simplex between these nodes is called a candidate $k$-simplex. Such as (2,3,7) in Fig.~\ref{fig1}(b), which is a candidate 3-simplex which contains \{(2),(3),(7),(2,3),(2,7),(3,7)\}. The process by which a candidate $k$-simplex evolves into a $k$-simplex over time is called a simplicial closure event\cite{benson2018simplicial}. Fig.~\ref{fig1}(d) shows a simplicial closure event of node 3,4 and 5. Over time, the three nodes tend to connect with each other
and form a candidate 3-simplex at $t_3$ timestamp, and finally form a 3-simplex at $t_5$ timestamp. In this paper, higher-order link prediction refers to the prediction of simplicial closure events in a simplicial network.

To better identify candidate $k$-simplices, we define open and closed cliques according to their status within the simplicial network and the corresponding skeleton network. {As shown in Fig.~\ref{fig1}(c), a skeleton network is composed of the nodes and edges extracted from the corresponding simplicial network. In a skeleton network, a fully connected subgraph with $k$ nodes is called a $k$-clique. And simplices in simplicial network can be projected as cliques in the skeleton network. But not all cliques in a skeleton network are projections of the simplices in its' corresponding simplicial network. This can also show that finding all cliques through a pairwise network(see Fig.~\ref{fig1}(c)) cannot derive the true higher order network(see Fig.~\ref{fig1}(b)).} Here, we define a closed $k$-clique as one who is  simultaneously contained within a simplex, and define an open $k$-clique as one who does not exist simultaneously within a simplex. Taking an example, the 3-clique (2, 3, 7) in Fig.~\ref{fig1}(c) is an open clique since it does not appear in the higher-order interactions depicted in Fig.~\ref{fig1}(a). On the other hand, the 3-clique (3, 5, 6) in Fig.~\ref{fig1}(c) corresponds to the 3-simplex shown in Fig.~\ref{fig1}(a) at the $t_2$ timestamp, indicating that it is a closed clique. It can be found that an open $k$-clique actually has all subsets of a $k$-simplex. In fact, an open $k$-clique is a candidate $k$-simplex. For instance, the open 3-clique (2,3,7), is also a  candidate 3-simplex. More obviously, a closed clique is a simplex in a higher-order network. Here, the skeleton network is not the focus of this paper, and it is mainly used to help identify candidate simplicies in simplicial networks.

\subsection{Higher-order link prediction problem}
Traditional link prediction typically involves predicting missing or false links between two nodes in a pairwise network, inferring connections between pairs of nodes. Higher-order link prediction, however, also forecasts interactions(simplices) involving multiple nodes in simplicial networks. The simplicial network is a higher-order network. A temporal simplicial network is a dynamic evolving higher-order network, and all interactions are arranged chronologically. Therefore, we can predict whether there will be a $k$-simplex (higher-order interaction) between $k$ nodes over time. More specifically, the higher-order link prediction in this paper means predicting simplicial closure events in simplicial networks.

Take $k$-simplices (interactions) prediction as an example. We first sort out all simplices by timestamp, with the top 80\% being the training set and the bottom 20\% being the test set. Next, we identify all candidate $k$-simplices in the training set. Finally, we take the candidate $k$-simplices which in the test set are $k$-simplices, as the positive samples. The remaining candidate $k$-simplices which are not closed in the test set, are used as the negative samples. Our aim is to predict which of the candidate $k$-simplices in the training set turn out to be $k$-simplices in the test set.

\subsection{Evaluation indicator}
The evaluation indicator used in this paper is the ratio of $PR-AUC$ to random baseline, and the greater the ratio, the better the prediction effect. $PR-AUC$ is the area under the Precision and Recall curves with different thresholds, and the random baseline refers to the proportion of candidate $k$-simplices in the training set transformed into $k$-simplices in the test set.

\begin{equation}
performance = \frac{\textit{PR-AUC}}{random\ baseline}.
\end{equation}
\subsection{Benchmarks}
Here, we select five indexes as benchmarks for comparison with the proposed method. The benchmarks include four higher-order extensions of high-performance pairwise algorithms and the current high-performance similarity metric based on the number of simplices. Each of these methods is described in detail below.  

\subsubsection{Higher-order extension of the methods based on high-performance pairwise algorithms}

In traditional link prediction task in pairwise networks, the indexes based on the local similarity of common neighbors always have outstanding performance. These indices measure the similarity between two nodes in a network based on the number of common neighbors they have. In this paper, the traditional common-neighbor-based algorithms are extended to higher-order networks to serve as the baseline algorithms for link prediction. This means we need to measure the similarity between $k$ nodes when predicting a $k$-simplex. To be specific, $k$-order common neighbor (KCN) index denotes the number of common neighbors of $k$ nodes, where $\Gamma(i)$ represents the set of neighbor nodes of node $i$. $K$-order Adamic-Adar (KAA) index represents the sum of the log of degrees of the common neighbors of $k$ nodes, where $k(x_i)$ denotes the degree of node $i$. $K$-order resource allocation (KRA) index represents the sum of the reciprocal of degrees of the common neighbors of $k$ nodes. And $k$-order preferential attachment (KPA) index represents the product of degrees of $k$ nodes.

\begin{equation}
s_{\left(x_{1}, \cdots, x_{k}\right)}^{KCN}=\left|\Gamma\left(x_{1}\right) \cap \Gamma\left(x_{2}\right) \cap \cdots \cap \Gamma\left(x_{k}\right)\right|,
\end{equation}

\begin{equation}
s_{\left(x_{1}, \cdots, x_{k}\right)}^{KAA}=\sum_{z \in \Gamma\left(x_{1}\right) \cap \cdots \cap \Gamma\left(x_{k}\right)} \frac{1}{\log k_{z}},
\end{equation}

\begin{equation}
s_{\left(x_{1}, \cdots, x_{k}\right)}^{KRA}=\sum_{z \in \Gamma\left(x_{1}\right) \cap \cdots \cap \Gamma\left(x_{k}\right)} \frac{1}{k_{z}},
\end{equation}

\begin{equation}
s_{\left(x_{1}, \cdots, x_{k}\right)}^{KPA}=k_{x_{1}} \cdot k_{x_{2}} \cdots \cdots k_{x_{k}}.
\end{equation}

\subsubsection{The algorithm based on the number of simplices}

An existing higher-order link prediction method takes into account the weights of edges in the skeleton graph of a simplicial network. It measures the similarity of higher-order interactions (simplices) based on the weights of all the edges in it. In the paper \cite{benson2018simplicial}, researchers define the weight $sw(x_i,x_j)$ of an edge $(i,j)$ as the number of interactions it contains. This method is called Simplex Weight (SW) in this paper. A $k$-simplex has $k(k-1)/2$ edges, and the average weight $sw(x_i,x_j)$ of these edges is taken as the similarity score in higher-order link prediction. There are different methods to calculate the average weight, so three indexes are derived, which are arithmetic average (SWA), geometric average (SWG), and harmonic average (SWH). These SW-based algorithms are algorithm is currently one of the best, with high prediction performance and the lowest computational complexity. Mathematically, they can be described as

\begin{equation}
s_{\left(x_{1}, \cdots, x_{k}\right)}^{SWA}=\left(\sum_{(x_i,x_j) \in k-\textit { simplex }} sw(x_i,x_j)\right) /\left(\frac{k(k-1)}{2}\right),
\end{equation}

\begin{equation}
s_{\left(x_{1}, \cdots, x_{k}\right)}^{SWG}=\left(\prod_{(x_i,x_j) \in k-\textit { simplex }} sw(x_i,x_j)\right)^{2 / k(k-1)} /\left(\frac{k(k-1)}{2}\right),
\end{equation}

\begin{equation}
s_{\left(x_{1}, \cdots, x_{k}\right)}^{SWH}=\left(\frac{k(k-1)}{2}\right) / \left(\sum_{(x_i, x_j) \in k-\textit { simplex }} sw(x_i, x_j)^{-1}\right),
\end{equation}
where $k$ denotes the order of the simplex that we want to predict, $sw(x_i,x_j)$ denotes the weight of edge $(x_i,x_j)$.

\section{Results}
\subsection{Data description}
In the paper, we use eight datasets for higher-order link prediction. Table ~\ref{table:T1} shows their statistics, and the specific description is as follows: 

Human contact data (\emph{contacting-high-school\cite{mastrandrea2015contact}, \emph{contacting-primary-school}\cite{stehle2011high}). The data is derived from interactive records of wearable sensors that were collected from both primary and high school students. The sensor records proximity-based contacts every 20 seconds. A simplex denotes there is an interaction between these individuals at each time interval.}

Ask Ubuntu thread data (\emph{threads-ask-ubuntu}). A node is a user, and a simplex is the set of questioners and answerers under the same question. The timestamp represents the question's posting time.

Ask Ubuntu tagging data (\emph{tags-ask-ubuntu}). The stack Exchange contains a series of Q \& A websites. Each website includes questions in different fields, and each question has multiple tags. Each tag is a node, tags related to the same question form a simplex, and each timestamp indicates that question's posting time.

U.S. National Drug Code data (\emph{NDC-classes}, \emph{NDC-substances}). The data includes drug class labels and drug constituent substances. In \emph{NDC-classes}, nodes are the class labels of drugs, and a simplex represents the class labels of a same drug. In \emph{NDC-substances}, each node refers to a kind of substance, and a simplex represents all substances in a drug. The timestamp in both datasets is the date a drug first hits the market.

Drug Abuse Warning Network (\emph{DAWN}) \cite{benson2018simplicial}. A national health warning system includes patients' pre-visit medication use records in U.S. hospital emergency rooms. Nodes represent drugs, and a simplex is the set of drugs a patient reports using before an emergency department visit.

Microsoft Academic Graph coauthorship data (\emph{coauth-MAG-History}). The Microsoft Academic Graph (MAG) is a knowledge graph of scholarly publications. This data is based on MAG's records in the field of history. A node is an author, and a simplex represents the authors of a scientific publication. The timestamp is the year of publication.

\subsection{The algorithm based on simplicial decomposition}
Each simplex contains multiple faces, which are themselves simplices of lower dimension. For example, a 4-simplex (1,2,3,4) contains the set of all faces $\{(1,2,3),(1,2,4),(2,3,4),(1,3,4),(1,2),(1,3),(1,4),(2,3),$\\
$(2,4),(3,4),(1),(2),(3),(4)\}$. Based on this idea, we propose a method called simplicial decomposition weight (SDW), which projects the higher-order information after simplicial decomposition as the weights to the edges in the skeleton networks corresponding to the simplicial network. The weight between nodes $x_i$ and $x_j$ is called $sdw(x_i,x_j)$. When predicting $k$-simplices and computing $sdw(x_i,x_j)$, we first identity the simplices containing nodes \emph{i} and \emph{j} in interaction data. Then we obtain the \emph{k}-faces of the set of simplices. We refer to the set of $k$-faces as the initial candidate sequence. Next, we add the $q$-faces of the initial candidate sequence ($q=k-1,...,2$) and update the candidate sequence. Note that all faces computed here are those containing nodes $i$ and $j$. Finally, the size of the candidate sequence is $sdw(x_i,x_j)$. Taking the example of calculating $sdw(3,4)$ when we will predict 3-simplices in fig 1. We first find the simplices containing nodes 3 and 4 in interaction data in Fig.~\ref{fig1}(a). The set of simplices is \{(1,2,3,4),(3,4,5)\}. Since the aim is to predict 3-simplices, which means $k = 3$ and $q = 2$, we obtain the 3-faces, i.e. \{(1,3,4),(2,3,4),(3,4,5)\}, as the initial candidate sequence. Next, we add the 2-faces of the initial candidate sequence is \{(3,4),(3,4),(3,4)\}. The final candidate sequence is \{(1,3,4),(2,3,4),(3,4,5),(3,4),(3,4),(3,4)\}. Thus $sdw(3,4)=6$. Note the faces in candidate sequence only need to contain nodes 3 and 4. We define the similarity of a simplex as the three averages of the simplicial decomposition weights of its edges, which are the arithmetic mean (SDWA), the geometric mean (SDWG), and the harmonic mean (SDWH). We refer to these three algorithms uniformly as SDW-based algorithms.

\begin{equation}
s_{\left(x_{1}, \cdots, x_{k}\right)}^{SDWA}=\left(\sum_{(x_i,x_j) \in k-\textit {simplex}} sdw(x_i,x_j)\right) /\left(\frac{k(k-1)}{2}\right),
\end{equation}

\begin{equation}
s_{\left(x_{1}, \cdots, x_{k}\right)}^{SDWG}=\left(\prod_{(x_i,x_j) \in k-\textit {simplex}} sdw(x_i,x_j)\right)^{2 / k(k-1)} /\left(\frac{k(k-1)}{2}\right),
\end{equation}

\begin{equation}
s_{\left(x_{1}, \cdots, x_{k}\right)}^{SDWH}=\left(\frac{k(k-1)}{2}\right) / \left(\sum_{(x_i, x_j) \in k-\textit {simplex}} sdw(x_i, x_j)^{-1}\right),
\end{equation}
where $k$ denotes the simplex' s order that we aim to predict, $sdw(x_i,x_j)$ denotes the simplicial decomposition weight of edge $(x_i,x_j)$.

\subsection{The algorithm based on open and closed cliques}
With the help of the simplicial network and the skeleton network, we can find all cliques and further divide them into opened cliques and closed cliques, which also provides new information about higher-order interactions. Based on this idea, we propose a method called closed ratio weight (CRW), to calculate the weight $crw(x_i,x_j)$ between node $x_i$ and node $x_j$. $crw(x_i,x_j)$ is the ratio of the number of closed cliques contain $(x_i,x_j)$ to the total number of cliques (both open and closed) containing $(x_i,x_j)$. In this study, when predicting $k$-simplices, we calculate the number of $2 \sim k$ order closed cliques, and the number of $3 \sim k$ order open cliques (there are no open 2-cliques in the network). Take (2, 3) in Fig.~\ref{fig1} as an example. When we predict 3-simplices, the closed cliques containing (2,3) are {(2,3), (1,2,3), (2,3,4)} and the open clique is {(2,3,7)}, so $crw(x_i,x_j)=3/4$. We define the similarity of a simplex as the three averages of the closed cliques' ratio weights of its edges, which are the arithmetic mean (CRWA), the geometric mean (CRWG), and the harmonic mean (CRWH). We refer to these three algorithms uniformly as CRW-based algorithms.

\begin{equation}
s_{\left(x_{1}, \cdots, x_{k}\right)}^{CRWA}=\left(\sum_{(x_i,x_j) \in k-\textit {simplex}} crw(x_i,x_j)\right) /\left(\frac{k(k-1)}{2}\right),
\end{equation}

\begin{equation}
s_{\left(x_{1}, \cdots, x_{k}\right)}^{CRWG}=\left(\prod_{(x_i,x_j) \in k-\textit {simplex}} crw(x_i,x_j)\right)^{2 / k(k-1)} /\left(\frac{k(k-1)}{2}\right),
\end{equation}

\begin{equation}
s_{\left(x_{1}, \cdots, x_{k}\right)}^{CRWH}=\left(\frac{k(k-1)}{2}\right) / \left(\sum_{(x_i, x_j) \in k-\textit {simplex}} crw(x_i, x_j)^{-1}\right),
\end{equation}
where $k$ denotes the simplex's order that we aim to predict, $crw(x_i,x_j)$ denotes the closed cliques ratio weight of edge $(x_i,x_j)$.

\subsection{Predicting 3-simplices in empirical simplicial networks}

In this section, we analyze the performance of the proposed SDW-based and CRW-based algorithms by comparing them with five high-performance local similarity algorithms in eight empirical simplicial networks. And the results are shown in the Table~\ref{table:T2}. In this experiment, the training set contains the first 80\% simplices, and the test set includes the remaining 20\% simplices. Table~\ref{table:T2} shows that  the indexes based on local higher-order information(i.e. SW-based, SDW-based, and CRW-based algorithms) are better than the higher-order extension of the traditional local indexes, which is consistent with the results of previous study \cite{benson2018simplicial}. Furthermore, the newly proposed SDW-based and CRW-based algorithmsperform even better than SW-based algorithms in five of eight networks (i.e. \emph{threads-ask-ubuntu}, \emph{tags-ask-ubuntu}, \emph{NDC-classes}, \emph{DAWN}, and \emph{coauth-MAG-History}). It verifies our original hypothesis that using more higher-order structure information can effectively improve prediction performance. In \emph{contact-high-school} network and \emph{contact-primary-school} network, although SDW-based algorithms and CRW-base algorithms may not be the best, but they are still among the top three, ranking second and third respectively. In terms of the SDW-based and CRW-based indexes themselves, the simplicial decomposition (SDW-based) method generally outperforms the closed ratio weight (CRW-based) method.

\subsection{Predicting 4-simplices in empirical simplicial networks}
In higher-order link prediction, only 3-order interactions are usually predicted, but there will be higher-order interactions in reality. Theoretically, searching simplices in a network is an NP-hard problem, making it challenging for higher-order link prediction. In this section, we try to predict 4-simplices in eight simplicial networks using the eight algorithms mentioned before.

As can be seen from Table~\ref{table:T3}, the prediction performance of the proposed indexes (SDW-based or CRW-based) get the best prediction performance in all networks except for \emph{NDC-substances} network. Notably, despite the increased difficulty in predicting 4-simplices, our proposed algorithms exhibit even better performance than when predicting 3-simplices, further widening the gap with other algorithms. Perhaps this is because our to SDW-based algorithms and CRW-based algorithms take more higher-order information into account when predicting 4-simplices, which is evident in the decomposition of more faces of each higher-order simplex.

\subsection{Robustness of higher-order link prediction}
\begin{figure*}[htb]
    \centering
    \includegraphics[width=1 \textwidth]{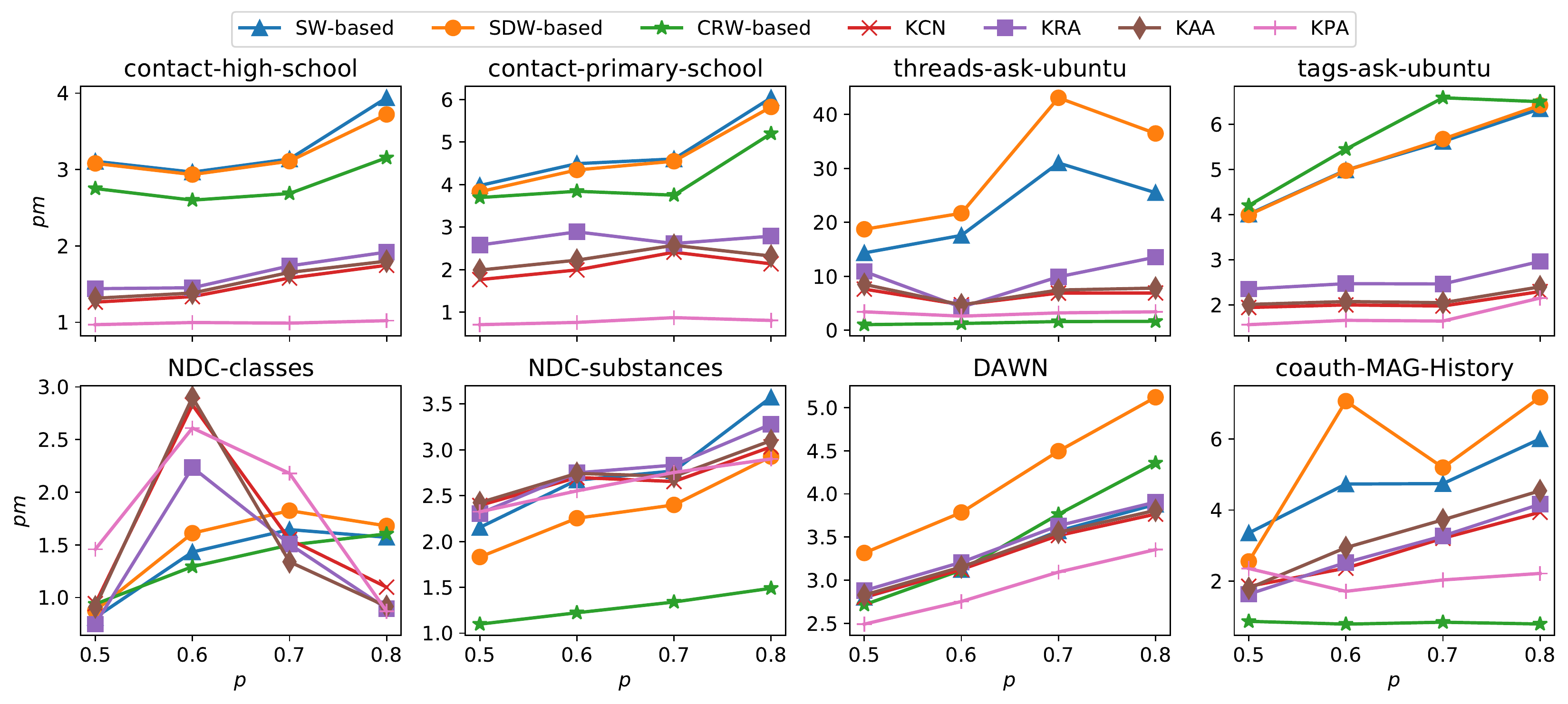}
    \caption{\textbf{(Color online)The robustness of predicting 3-simplices in eight simplicial networks.} We tested the performance of seven algorithms with different sizes of observed simplices in eight networks. $p$ on the x-axis represents the proportion of the training set. The y-axis represents the predicted performance $pm$.}
    \label{fig:3 order performance}
\end{figure*}

The criteria for a link prediction algorithm are: first, it performs well when most of the structures can be observed; second, it can also have high performance when only a few structures are known. To this end, we test the performance of seven algorithms with different sizes of observed simplices in eight networks. The size $p$ of the training set was selected from 50\% to 80\% of all data, with an interval of 10\%. Fig.~\ref{fig:3 order performance} shows the performance of different sizes of train sets in eight networks when predicting the 3-simplices. These results show that the proposed algorithms have the best performance in most networks corresponding to the training sets with different sizes, and the performance gets better as the size of the training set increases. It indicates that the more data you observe, the more information you know, the fewer samples you need to predict, and the better the prediction performance will be. But surprisingly, in \emph{NDC-classes}, KCN algorithm has the best prediction performance when the training set proportion is 60\%, and it outperforms all other algorithms.

\begin{figure*}[htb]
    \centering
    \includegraphics[width=1\textwidth]{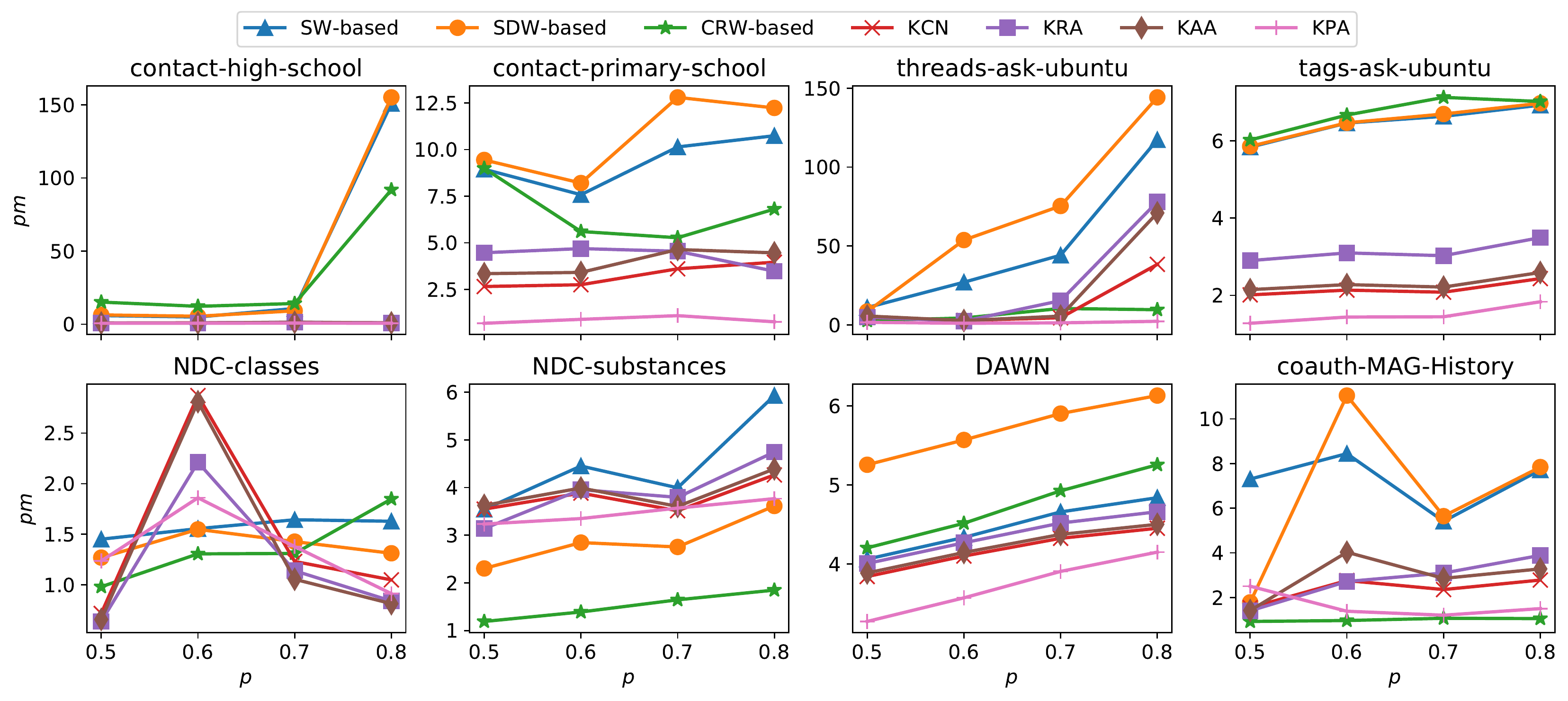}
    \caption{\textbf{(Color online)The robustness of predicting 4-simplices in eight simplicial networks.} We tested the performance of seven algorithms with different sizes of observed simplices in eight networks. $p$ on the x-axis represents the proportion of the training set. The y-axis represents the predicted performance $pm$.}
    \label{fig:4 order performance}
\end{figure*}

Fig.~\ref{fig:4 order performance} shows the performance of the seven algorithms with different sizes of training sets in eight networks when predicting 4-simplices. In general, it is more difficult to predict higher-order simplex. However, our proposed algorithms show a greater advantage when predicting the 4-simplex except in \emph{NDC-classes} and \emph{NDC-substances}. These results show that the proposed algorithms have the best performance in most networks corresponding to different sizes of training sets. And the performance gets better as the size of the training set increases. But surprisingly, a few networks can be predicted more accurately when less information is known, SW-based algorithm has the best prediction performance when the training set proportion is 60\% in \emph{coauth-MAG-History}; KAA algorithm has the best prediction performance when the training set proportion is 50\% in \emph{NDC-classes}. 

\subsection{Computational complexity analysis}

For the three algorithms, the average weights are calculated in the same way using arithmetic average, geometric average, and harmonic average. Therefore, we only need to consider the computational complexity of the three algorithms in computing the weights of the edges, i.e., $sw(x_i, x_j)$, $sdw(x_i, x_j)$, and $crw(x_i, x_j)$. Simplex weight (SW) defines the weight $sw(x_i, x_j)$ of an edge $(x_i, x_j)$ as the number of higher-order interactions it contains. If there are $m$ timestamped simplices, and we need to determine whether nodes $i$ and $j$ are contained in each simplex, and the algorithm complexity is $O(2m)$. Simplicial decomposition weight (SDW) decomposes simplex more than $k$-order based on SW algorithm. There are $m$ simplices that need to be decomposed in the worst case. Decomposing a $g$-simplex from order 2 to $k$ requires $O(\sum_{q=2}^{k} C_{g}^{q})$. We restrict the maximum order of the simplex to 25, then the maximum value of $g$ is 25 and the maximum value of $k$ is 4. In the worst case, the computational complexity of the SDW algorithm is $O(2m+m\sum_{q=2}^{4} C_{25}^{q}) \approx O(2m)$, which also grows linearly. The closed ratio weight (CRW) algorithm calculates the number of open cliques of order $3$ to $k$ containing edge $(i,j)$ based on the SW algorithm. We find all $k$-order open cliques by enumerating the $k$-cliques. According to the paper\cite{zhang2005genome}, the computational complexity of enumerating $k$-cliques is $O((n-k)^2 N[k]+nM[k]) \approx O(n^2)$, where $n$ is the number of nodes, and $N[k]$ and $M[k]$ are constants related to different network structures. Therefore, the computational complexity of the CRW algorithm is $O(2m+n^2)$.

\section{Conclusion}
\label{sec:conclusion}
In summary, we explore higher-order link prediction in simplicial networks via local information. Unlike the traditional link prediction, we predict higher-order links beyond pairwise links. We propose two novel indices to measure the similarity of higher-order links, SDW and CRW. SDW is based on simplex decomposition. Each simplex contains not only itself but also each of its lower-order faces. Therefore, we consider not only the simplices but also the faces during the decomposition of these simplices. This index further enhances the contribution of the higher-order simplices. CRW takes into account the state (closed or open) of the cliques in which the edge is located. The higher the value, the more the tested sample tends to the closed state. Both of these proposed indexes provide the more details of local higher-order information to measure the similarity of nodes in the higher-order link prediction.

We evaluate the proposed algorithms SDW and CRW and other five benchmark algorithms in eight networks. The results show that the proposed algorithms outperforms others in most networks when predicting 3-simplices and they have even better performance in 4-simplices prediction. Besides, we evaluate the algorithms' robustness by testing their link prediction performance with different sizes of training sets. And the results show that in most networks with varying sizes of observed data, the prediction performance of the proposed algorithms gets better with the increase of observed data. However, there are exceptions. For example, when predicting 4-simplices in the \emph{coauth-MAG-History} network, SDW algorithm can achieve better performance with only a small amount of data than with more data.

For future work, we will try to find novel methods to predict higher-order interactions. In addition, it will be attractive to explore the relationship between prediction algorithms, the algorithms corresponding to different types of higher-order networks, and higher-order predictability. 

\begin{acknowledgments}
The authors acknowledge the Special Project for the Central Guidance on Local Science and Technology Development of Sichuan Province (Grant No. 2021ZYD0029), the STI 2030—Major Projects (Grant No. 2022ZD0211400), the National Natural Science Foundation of China (Grant No. T2293771) and the New Cornerstone Science Foundation through the XPLORER PRIZE.
\end{acknowledgments}

\section*{Data Availability Statement}

The networks data that support the findings of this study are available through the corresponding references\cite{benson2018simplicial}.

\section*{Conflict of Interest Statement}

The authors declare that the research was conducted in the absence of any commercial or financial relationships that could be construed as a potential conflict of interest.

\section*{Author Contributions}

B.L. and L.L. conceived the idea and designed the research. B.L. performed the experiments and wrote the manuscript.  R.Y. drew the figures and made the tables. L.L. edited this paper and supervised this work. All authors contributed to the discussion on the data and revision of the manuscript.

\bibliography{refs}

\clearpage
\begin{table*}[htb]
\caption{\textbf{Statistics of eight networks.} Statistics include the number of nodes, the number of edges in skeleton network, the number of simplices.}
\label{table:T1}
\begin{ruledtabular}
\begin{tabular}{cccc}
Dataset & Nodes & Edges & Simplices   \\ \hline
contact-high-school    & 327             & 5818            & 172035                \\
contact-primary-school & 242             & 8317            & 106879                \\
threads-ask-ubuntu     & 125602          & 187157          & 192947               \\
tags-ask-ubuntu        & 3029            & 132703          & 271233                \\
NDC-classes            & 1161            & 6222            & 49724                 \\
NDC-substances         & 5311            & 88268           & 112405                \\
DAWN                   & 2558            & 122963          & 2272433               \\
coauth-MAG-History     & 1014734         & 1156914         & 1812511               \\ 
\end{tabular}
\end{ruledtabular}
\end{table*}

\begin{table*}[htb]
\caption{\textbf{The performance of predicting 3-simplices in eight simplicial networks.} Among the SW-based, SDW-based and CRW-based algorithms, each of them has three methods for averaging weights. The best-performing method for each algorithm is listed alongside its corresponding performance value.}
\label{table:T2}
\begin{ruledtabular}
\begin{tabular}{cccccccc}
Datasets  & SW-based    & KCN     & KRA    & KAA    & KPA    & SDW-based   & CRW-based   \\ \hline
contact-high-school    & \textbf{3.939(WG)}    & 1.746   & 1.919 & 1.8 & 1.021  & 3.721(WG)  & 3.154(WA) \\
contact-primary-school & \textbf{6.04(WG)}  & 2.137  & 2.788 & 2.321 & 0.807  & 5.825(WG)  & 5.199(WA) \\
threads-ask-ubuntu     & 25.499(WG) & 6.904 & 13.547 & 7.806 & 3.411  & \textbf{36.482(WG)} & 1.642(WH) \\
tags-ask-ubuntu        & 6.344(WG)  & 2.297 &  2.965  & 2.402   & 2.155  & 6.425(WG)  & \textbf{6.504(WA)} \\
NDC-classes            & 1.572(WG)  & 1.098  & 0.895 & 0.915 & 0.869   & \textbf{1.68(WG)}  & 1.604(WA) \\
NDC-substances         & \textbf{3.573(WG)}  & 3.032  & 3.279 & 3.1 & 2.9  & 2.926(WG)  & 1.49(WA) \\
DAWN                   & 3.879(WH)   & 3.766  & 3.905 & 3.809 & 3.355  & \textbf{5.119(WG)}  & 4.358(WG) \\
coauth-MAG-History     & 6.002(WG)   & 3.937  & 4.163 & 4.542 & 2.216  & \textbf{7.174(WA)}  & 0.793(WA) \\
\end{tabular}
\end{ruledtabular}
\end{table*}

\begin{table*}[htb]
\caption{\textbf{The performance of predicting 4-simplices in eight simplicial networks.} Among the SW-based, SDW-based and CRW-based algorithms, each of them has three methods for averaging weights. The best-performing method for each algorithm is listed alongside its corresponding performance value.}
\label{table:T3}
\begin{ruledtabular}
\begin{tabular}{cccccccc}
Datasets  & SW-based    & KCN     & KRA    & KAA    & KPA   & SDW-based   & CRW-based    \\ \hline
contact-high-school    & 150.872(WH)  & 0.92 & 0.903 & 0.89  &  0.677  & \textbf{155.09(WH)}  & 91.918(WH)  \\ 
contact-primary-school & 10.743(WH)  & 3.962 & 3.479 & 4.463 & 0.766  & \textbf{12.234(WH)}  & 6.81(WG) \\
threads-ask-ubuntu     & 117.34(WG) & 38.325 & 78.034 & 70.945 & 2.219 & \textbf{144.199(WH)} & 9.57(WH) \\
tags-ask-ubuntu        & 6.927(WG)  &  2.434 & 3.495 & 2.592 & 1.833  & 6.976(WG)  & \textbf{7.022(WG)} \\ 
NDC-classes            & 1.629(WA)  & 1.05 & 0.841 & 0.815 & 0.914  & 1.313(WG)  & \textbf{1.849(WA)}  \\ 
NDC-substances         & \textbf{5.929(WH)}  &  4.264 & 4.745 & 4.39 & 3.767  & 3.613(WG)  & 1.848(WH) \\
DAWN                  & 4.839(WH)   & 4.452 & 4.661 & 4.502 & 4.15  & \textbf{6.132(WG)}  & 5.256(WG)  \\
coauth-MAG-History     & 7.712(WA)   & 2.779 & 3.885 & 3.28 & 1.5 & \textbf{7.843(WA)}  & 1.052(WA) \\
\end{tabular}
\end{ruledtabular}
\end{table*}
\end{document}